\documentclass[12pt]{article}
\usepackage{epsfig}
\usepackage{hyperref}
\usepackage{amssymb}
\usepackage{graphics} 
\usepackage{amsbsy}
\usepackage{graphicx}
\usepackage[intlimits]{amsmath}
\usepackage{amsthm}
\usepackage{amssymb}
\usepackage{amstext}
\usepackage{verbatim}
\usepackage{mathrsfs}
\usepackage{latexsym}
\usepackage{float}
\usepackage{subfig}
\usepackage{wasysym}
\usepackage{fancyhdr}
\usepackage[table]{xcolor}
\usepackage{supertabular}
\usepackage{longtable}
\usepackage{setspace}
\usepackage{multirow}
\usepackage{lineno}
\usepackage{subfig}
\usepackage{indentfirst}
\usepackage{authblk}

\hypersetup{
 colorlinks,
  citecolor = black,
  filecolor = black,
  linkcolor = black,
 urlcolor = black,
  linktocpage
}

\usepackage[top=1in, bottom=1.in, left=1in, right=1in]{geometry}
\begin{document}

\title{\vspace*{-0.5in}
{\bf A Whitepaper on SoLID} \\
({\bf So}lenoidal {\bf L}arge {\bf I}ntensity {\bf D}evice) \\
}
\author{J.P. Chen$^{1}$}
\author{H. Gao$^{2}$} 
\author{T.K. Hemmick$^{3}$}
\author{Z.-E. Meziani$^{4}$} 
\author{P.A. Souder$^{5}$}
\author{the SoLID Collaboration}

\affil{$^{1}$Thomas Jefferson National Accelerator Facility, Newport News, VA, USA}
\affil{$^{2}$Triangle Universities Nuclear Laboratory and Department of Physics, Duke University, Durham, NC, USA}
\affil{$^{3}$Stony Brook University, Stony Brook, NY, USA}
\affil{$^{4}$Temple University, Philadelphia, PA, USA}
\affil{$^{5}$Syracuse University, Syracuse, NY, USA}

\date{\today}
\maketitle

\begin{figure}[!ht]
\begin{center}
\includegraphics[width=0.65\textwidth, angle = 0]{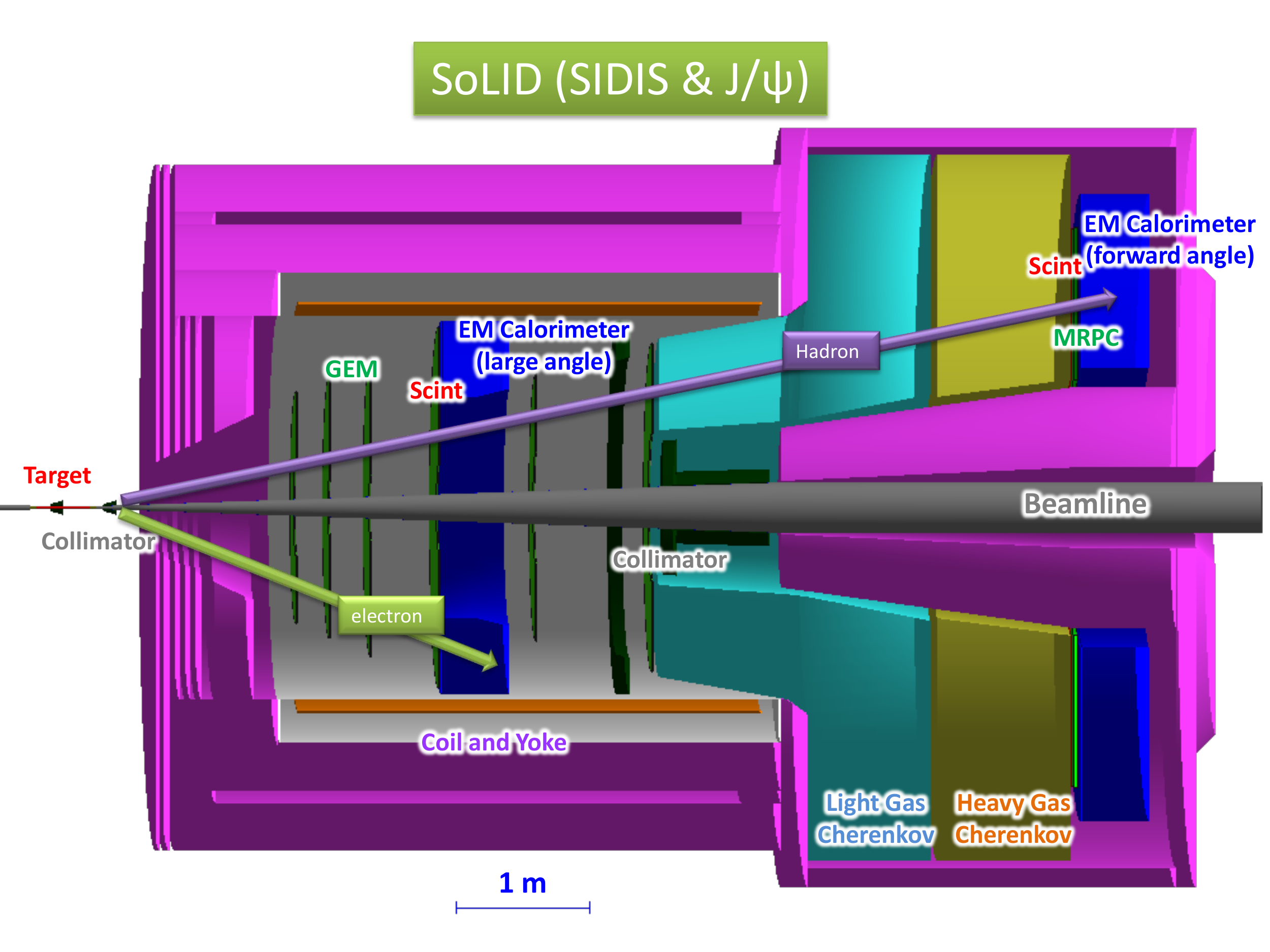}
\vspace{0.1 cm}
\end{center}
\label{fig:PV1}
\end{figure}


\newpage
\section{Introduction}
At the time of this writing, the highly anticipated 12 GeV upgrade of
the CEBAF accelerator is nearing completion and its rich physics
program is about to commence in earnest. 

The 12 GeV upgrade is accompanied by new detector upgrades that
include CLAS12, SHMS, and an entirely new experimental hall 
featuring searches for QCD exotic states (GlueX).  Although the
currently envisioned program includes both high rate capability and
large acceptance devices, there does not exist a single device capable
of handling high luminosity ($10^{36}-10^{39}$cm$^{-2}$s$^{-1}$) over a large 
acceptance.  {\bf The capabilities of the 12 GeV upgrade will not have been fully
  exploited unless a large acceptance high luminosity device is
  constructed.}  The SoLID ({\bf So}lenoidal {\bf L}arge {\bf
  I}ntensity {\bf D}etector) program is designed to fulfill this need.

SoLID is made possible by developments in both detector technology and
simulation accuracy and detail that were not available in the early
stages of the 12 GeV program planning.  The spectrometer is designed
with a unique capability of reconfiguration to optimize capabilities
for either Parity-Violating Deep Inelastic Scattering (PVDIS) or
Semi-Inclusive Deep Inelastic Scattering (SIDIS) /threshold production
of the J/$\psi$ meson.  Already four experiments with an ``A'' rating and
one with an ``A-'' rating have been approved for SoLID by the 
Jefferson Lab Program Advisory Committee. Figure 1 shows 
the two SoLID configurations for these experiments. The
collaboration has grown to include more than 200 collaborators at over 
50 institutions from 11 nations. A conceptual design of SoLID base equipment
has been fully developed and has gone through numerous internal discussions and
informal reviews in the last few years. The current status can be found in the SoLID 
pCDR~\cite{pCDR} document.

In the sections that follow, we detail the rich physics program that
can only be realized by the construction of SoLID at Jefferson Lab followed by an overview of the SoLID instrumentation and its current status.

\begin{figure}[!ht]
\begin{center}
\includegraphics[width=0.45\textwidth, angle = 0]{SoLID_setup_SIDIS}
\includegraphics[width=0.45\textwidth, angle = 0]{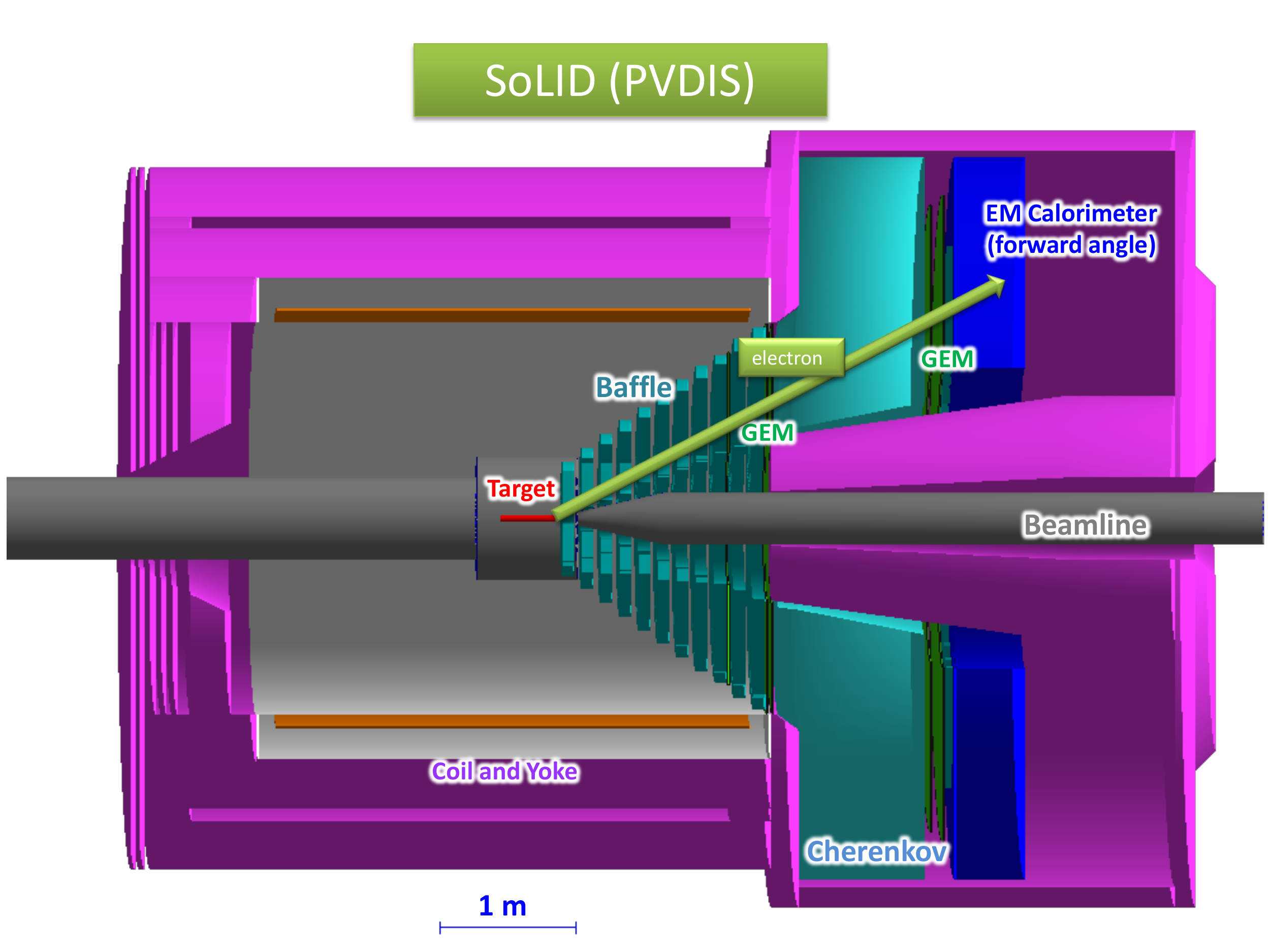}
\caption{Left panel: SoLID apparatus for the SIDIS and the J/Psi program. 
Right panel: SoLID apparatus for the PVDIS program.}
\label{fig:PV1}
\end{center}
\end{figure}

\section{Transverse Spin and Three-Dimensional momentum Structure of the nucleon from SIDIS}

Understanding the internal structure of nucleon and nucleus in terms of 
quarks and gluons, the fundamental degrees of freedom of Quantum Chromodynamics (QCD), has been, and still is, the frontier of subatomic physics research.
 In recent years, the hadronic physics community has
advanced its investigation of partonic structure of hadrons 
beyond the one-dimensional parton distribution functions (PDFs) by exploring the  motion and spatial distributions of the partons in the direction perpendicular to
the momentum of the parent hadron.
Such efforts are closely connected to the study and extraction of two new 
types of parton distribution functions concerning the tomography of 
the nucleon. They are the transverse-momentum-dependent parton distributions (TMDs), and the generalized
parton distributions (GPDs), providing new information about the rich dynamics of QCD. 

\begin{figure}[!ht]
\begin{center}
\includegraphics[width=0.55\textwidth, angle = 0]{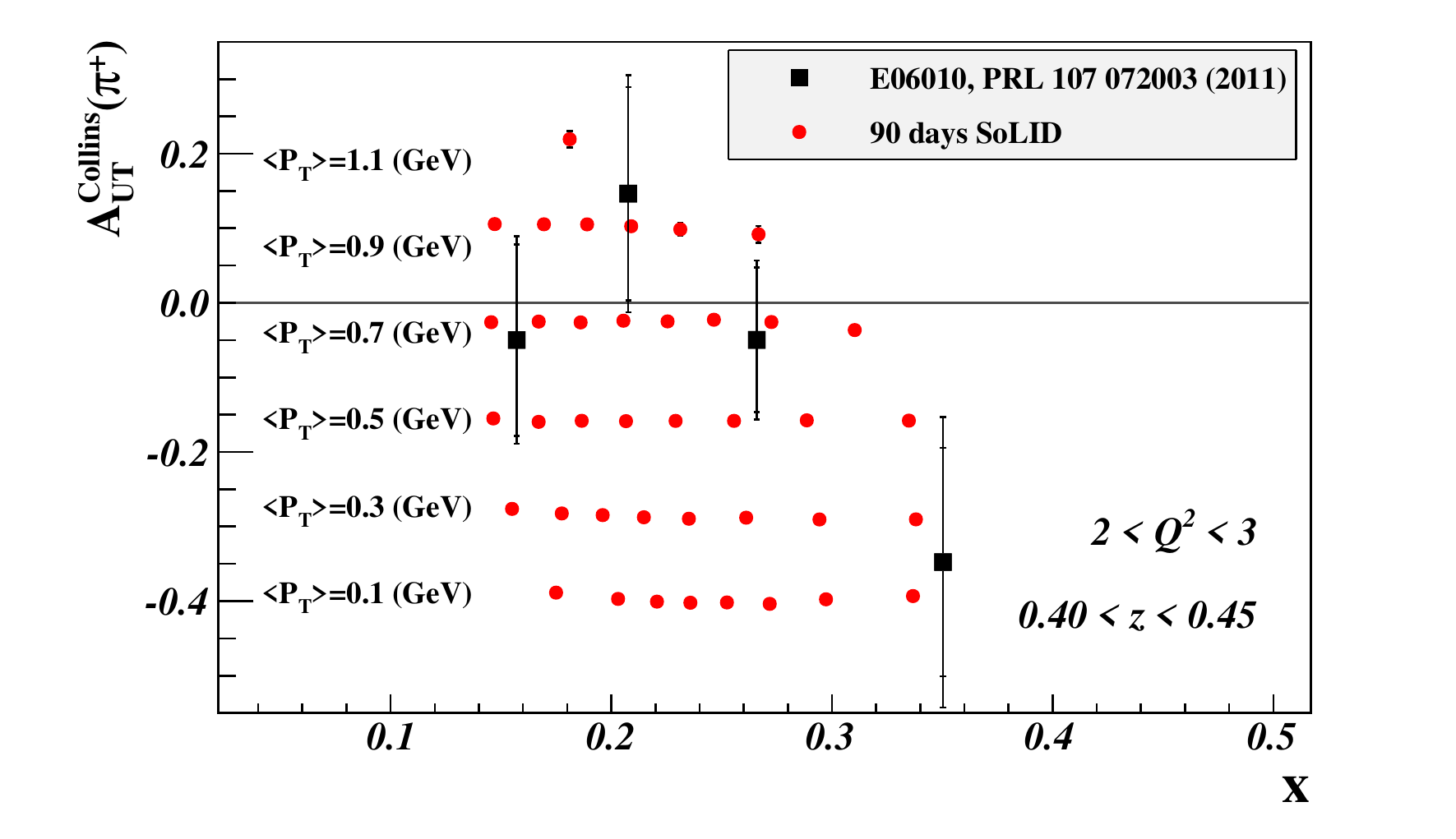}
\includegraphics[width=0.55\textwidth, angle = 0]{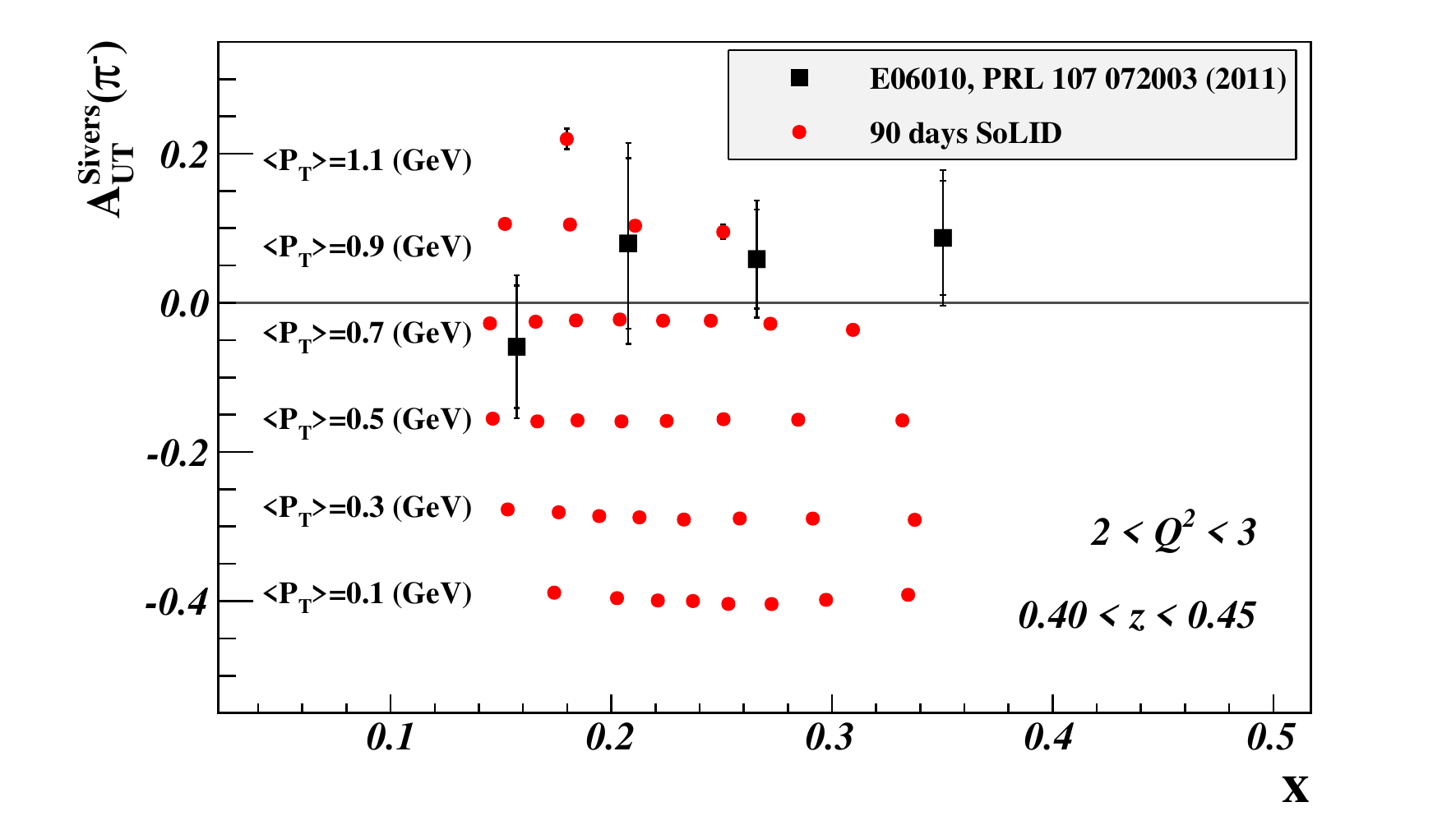}
\caption{Top panel: The projected Jefferson Lab 12-GeV SoLID measurement of the Collins asymmetry for $\pi^+$ from a transversely polarized $^3$He target for a $z$ bin of 0.4 to 0.45, and a $Q^2$ bin of 2 to 3 (GeV/c)$^2$ as a function of 
$\pi^+$ transverse momentum, and Bj\"{o}rken $x$. Also shown are the published result from the 6 GeV experiment~\cite{xqian}. 
Bottom panel: the projected measurement for the Sivers asymmetry for $\pi^-$ together with the published results from ~\cite{xqian}.}
\label{fig:SIDIS-projection}
\end{center}
\end{figure}

\begin{figure}[!ht]
\begin{center}
\includegraphics[width=0.55\textwidth, angle = 0]{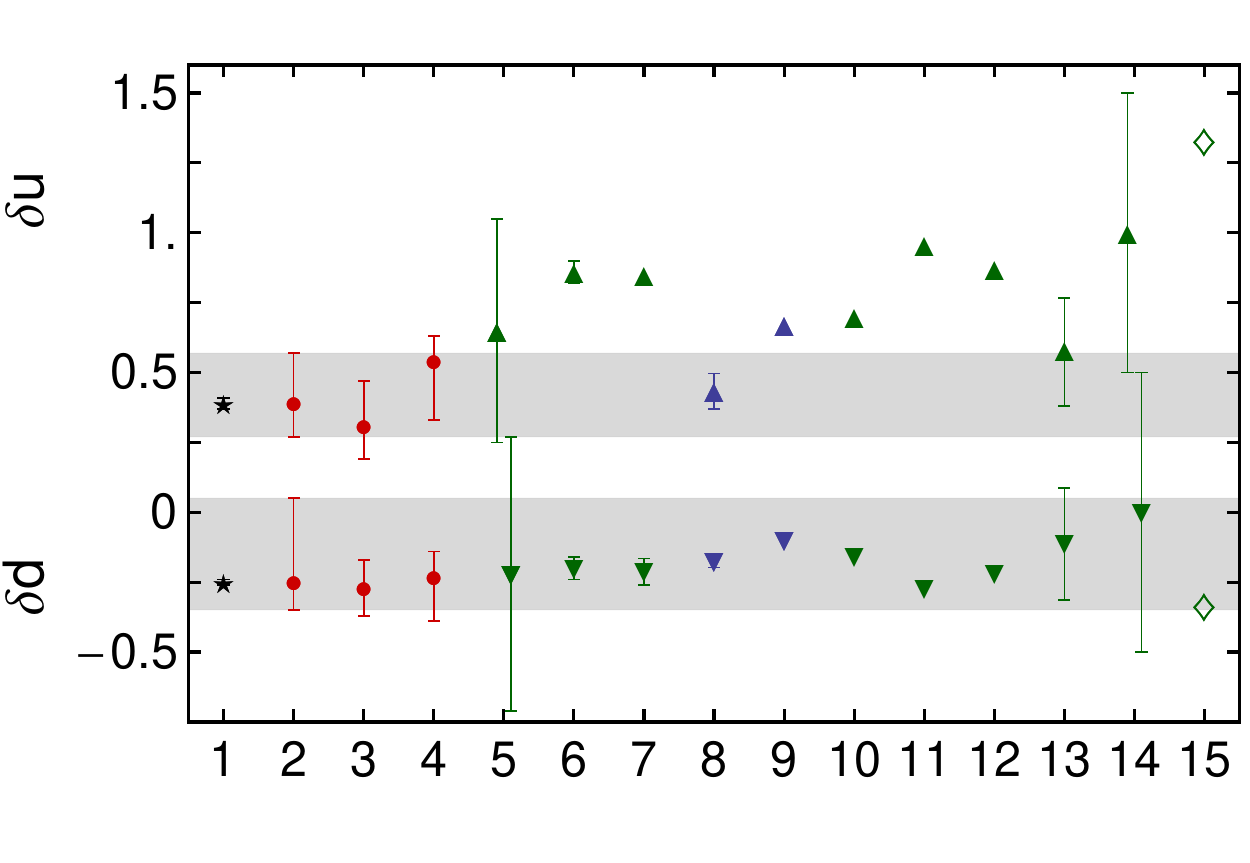}
\includegraphics[width=0.55\textwidth, angle = 0]{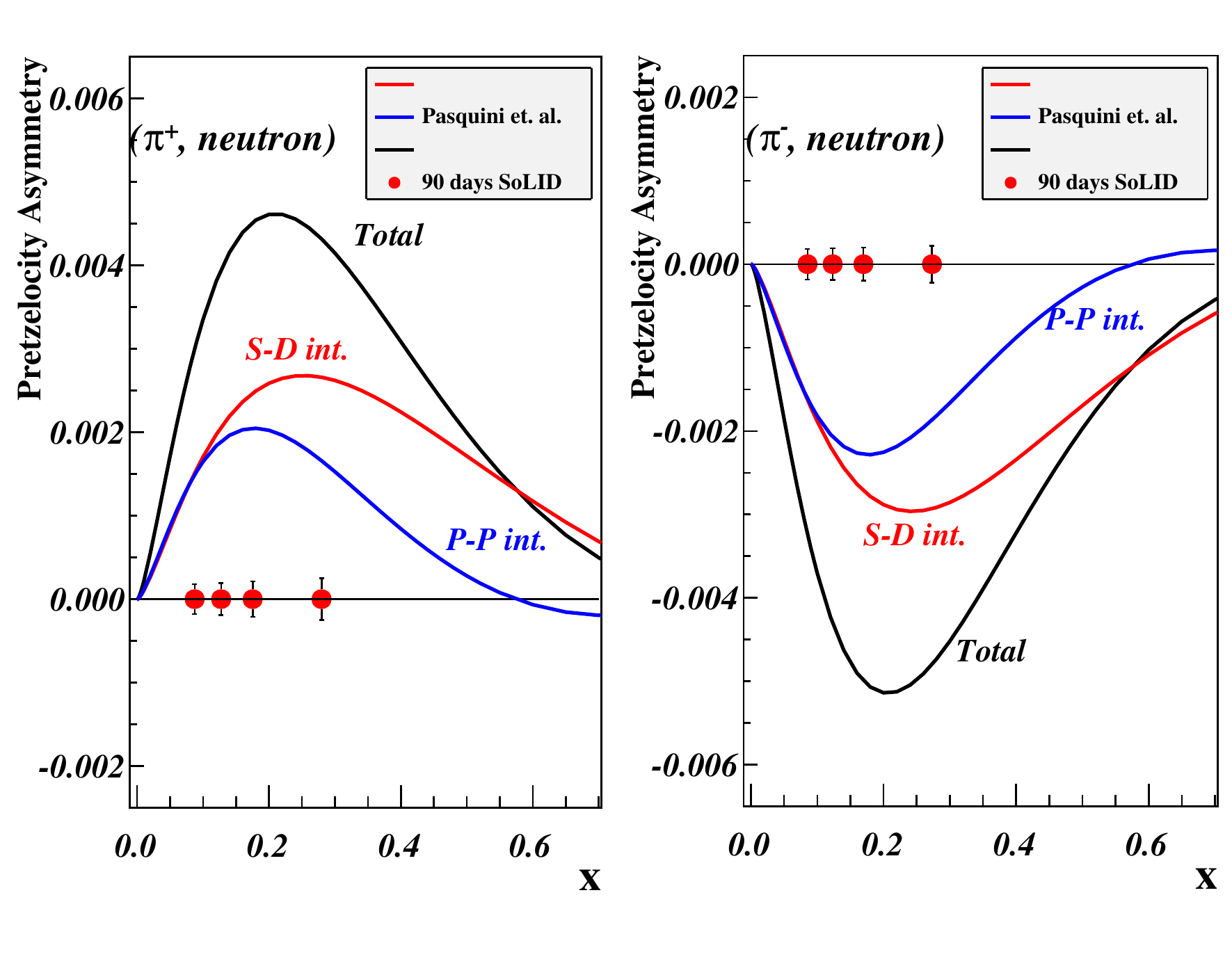}
\caption{Top panel: The projected Jefferson Lab 12-GeV SoLID determination of the tensor charge of u and d quark (black points) 
together with model dependent extractions of these quantities based on the existing data. Also shown are predictions from lattice QCD, Dyson-Schwinger Equations, and various models. 
Bottom panel: The projected pretzelocity asymmetry measurement from a transversely polarized $^3$He target (``neutron'') together with predictions from Boffi and Pasquini {\it et al.}~\cite{pasquini2009} at a $Q^2$ value of 2.5 GeV$^2$.}
\label{fig:tensor-charge}
\end{center}
\end{figure}

Knowledge of TMDs is essential to unfold the full momentum and spin structure of the nucleon.  The TMDs represent the confined motion of  
partons inside the nucleon and allow reconstruction of the
nucleon structure in three-dimensional momentum space, thereby leading to 
exploration and study of new structure, new dynamics, and new phenomena. 
 Most TMDs and related observables are due to couplings of the
transverse momentum of quarks with the spin of the nucleon (or the quark). 
Spin-orbit correlations in QCD similar to those in hydrogen atoms, 
can therefore be studied.
 Among the eight leading-twist TMDs, three survive the integration over the transverse momenta of the quarks, and they are the unpolarized and the
longitudinally polarized (helicity) TMDs, and the transversely polarized quark distribution function (transversity). Among these three, the transversity is the least known. Its lowest moment defines the tensor charge, a fundamental quantity related to the spin of the nucleon, therefore providing an excellent testing ground for lattice QCD predictions. 
 The remaining five leading-twist TMDs would vanish in the absence
of parton orbital angular momentum, therefore, they provide 
quantitative information about the orbital angular momentum (OAM) of the 
partons inside the nucleon.  
The origin of some TMDs and related spin asymmetries, at the partonic level, depends on fundamental properties of QCD, such as its color gauge invariance, which lead to predictions that can be tested experimentally. 

The Jefferson Lab 12 GeV energy upgrade and the SoLID detector with the combination of a large acceptance and high luminosity 
provides a golden opportunity to perform precision measurements of both the single and double spin asymmetries in semi-inclusive deep inelastic scattering, allowing multi-dimensional binning in all relevant kinematic quantities,   
 from polarized $^3$He (``neutron'') and polarized proton targets to 
extract TMDs with unprecedented precision and flavor separation in the valence quark region. Fig.~\ref{fig:SIDIS-projection} is an illustration showing the 
high precision of the projected measurement from a transversely polarized $^3$He target as a function of Bj\"{o}rken $x$, and pion transverse momentum for an example $z$, and $Q^2$ bin, among a total number of 1400 projected data points. Also shown are our published results from the 6 GeV experiment~\cite{xqian} for comparison.     
Such high precision will provide benchmark tests of dynamical lattice QCD predictions for the tensor charge of the u and d quark, provide quantitative information about quark OAM inside the nucleon, and uncover crucial information 
about the dynamics of QCD.  Furthermore, the SoLID dihadron program, which will take data parasitically with the SIDIS program, will complement the SIDIS measurements nicely, particularly in the extraction of the transversity TMD. 

The top panel of Fig.~\ref{fig:tensor-charge} is a compilation of our current, model-dependent 
knowledge about the u and d quark tensor charges determined from analyses~\cite{Anselmino2013,Anselmino2009,Bacchetta2013} of existing data (shown as points 2-5), the projected results from the JLab SoLID program within the same model~\cite{Prokudin2014} (point 1), together with predictions based on lattice QCD~\cite{Alexandrou2014, Gockeler2005} (points 6, and 7), Dyson-Schwinger equations~\cite{Pitschmann2014, Hecht2001} (points 8 and 9), and from various models~\cite{Cloet2008,Pasquini2007,Wakamatsu2007,Gamberg2001,He1995} (points 10 - 15).
The model dependent uncertainty in the latest extraction~\cite{Anselmino2013} is shown as a grey band. 
In the bottom panel of Fig.~\ref{fig:tensor-charge}, we show as an example, the projected measurements of the pretzelocity asymmetry on the neutron ($^3He$) for both $\pi^+$ and $\pi^-$ together with predictions from S. Boffi {\it et al.}~\cite{pasquini2009}, where contributions from interference of the $S$-$D$, and $P$-$P$ orbital angular momentum states are shown together with the total.   
It is clear that the projected high precision results from SoLID at 12-GeV JLab will provide powerful tests of LQCD predictions, and much needed quantitative information about quark OAM inside the nucleon.  


\section{Parity Violation in Deep-Inelastic Scattering}
In the late 70's, Prescott {\it et al.}~\cite{Prescott:1978tm, Prescott:1979dh} showed that the weak neutral current violates parity by measuring the non-zero asymmetry 
$A_{LR}=(\sigma_L-\sigma_R)/(\sigma_L+\sigma_R)$ for polarized electron-deuterium deep inelastic scattering.  The experiment also set limits on the dependence of $A_{LR}$  on the variable $y\equiv (E-E')/E$, which ruled out models invented to explain the negative results of the early atom parity violation (APV) experiments.  Subsequent to these publications, Glashow, Salam, and Weinberg were awarded the Nobel prize for electroweak unification.

Parity violating electron scattering (PVES) from deuterium in the DIS region at large Bjorken $x$ is an attractive reaction for searching for new physics since there the $A_{LR}$ is approximately independent of hadron structure, and the remaining QCD effects, which are of great interest in themselves, can be isolated by their kinematic dependence.  The asymmetry has the form $A/Q^2=-[a_1+a_3f(y)]$, where $f(y)\approx[1-(1-y)^2]/[1+(1-y)^2]$, $a_1\propto 2 C_{1u}-C_{1d}$ and $a_3\propto 2 C_{2u}-C_{2d}$. $C_{1q}(C_{2q})$ are four-Fermi coupling constants with axial(vector) electron currents and vector(axial) quark currents.  At large $y$, $A_{LR}$ is sensitive to $C_{2q}$, which cannot be studied in low energy reactions because of large and uncertain radiative corrections.
Experiments with atomic parity violation and PVES experiments including 
Qweak~\cite{Androic:2013rhu} have yielded precise measurements of $C_{1q}$.  The Jefferson Lab PVDIS collaboration~\cite{Wang:2014bba} has recently published in Nature a new experimental result,   
$2C_{2u}-C_{2d}=-0.145\pm0.068$. This is the first measurement sufficiently sensitive to show that $C_{2q}$ are non-zero as predicted by the Standard Model (SM).

\begin{figure}[!ht]
   \begin{center}
       \includegraphics[type=pdf, 
ext=.pdf,read=.pdf,width=0.53\textwidth]{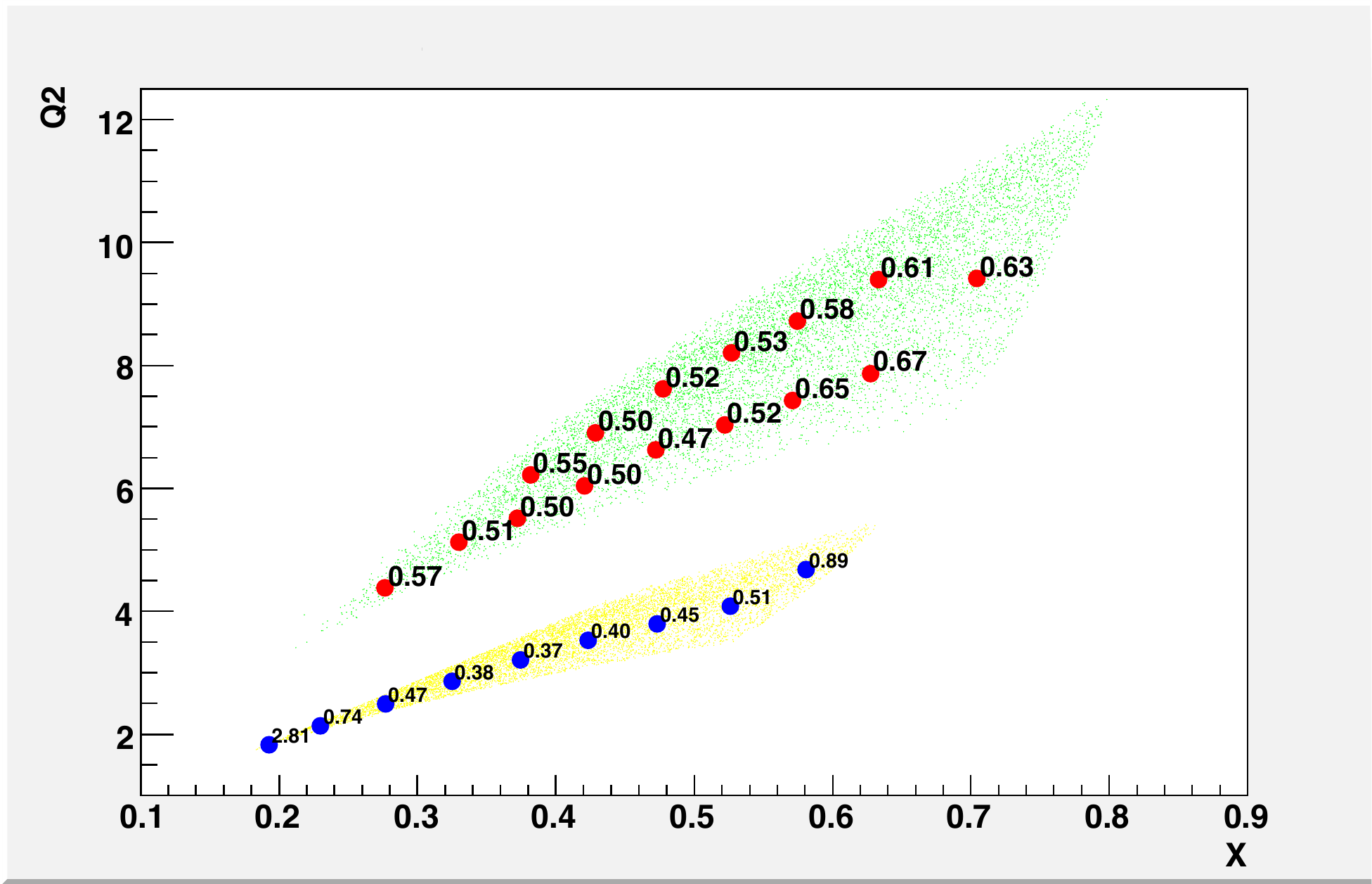}
       \includegraphics[type=pdf, 
ext=.pdf,read=.pdf,width=0.41\textwidth]{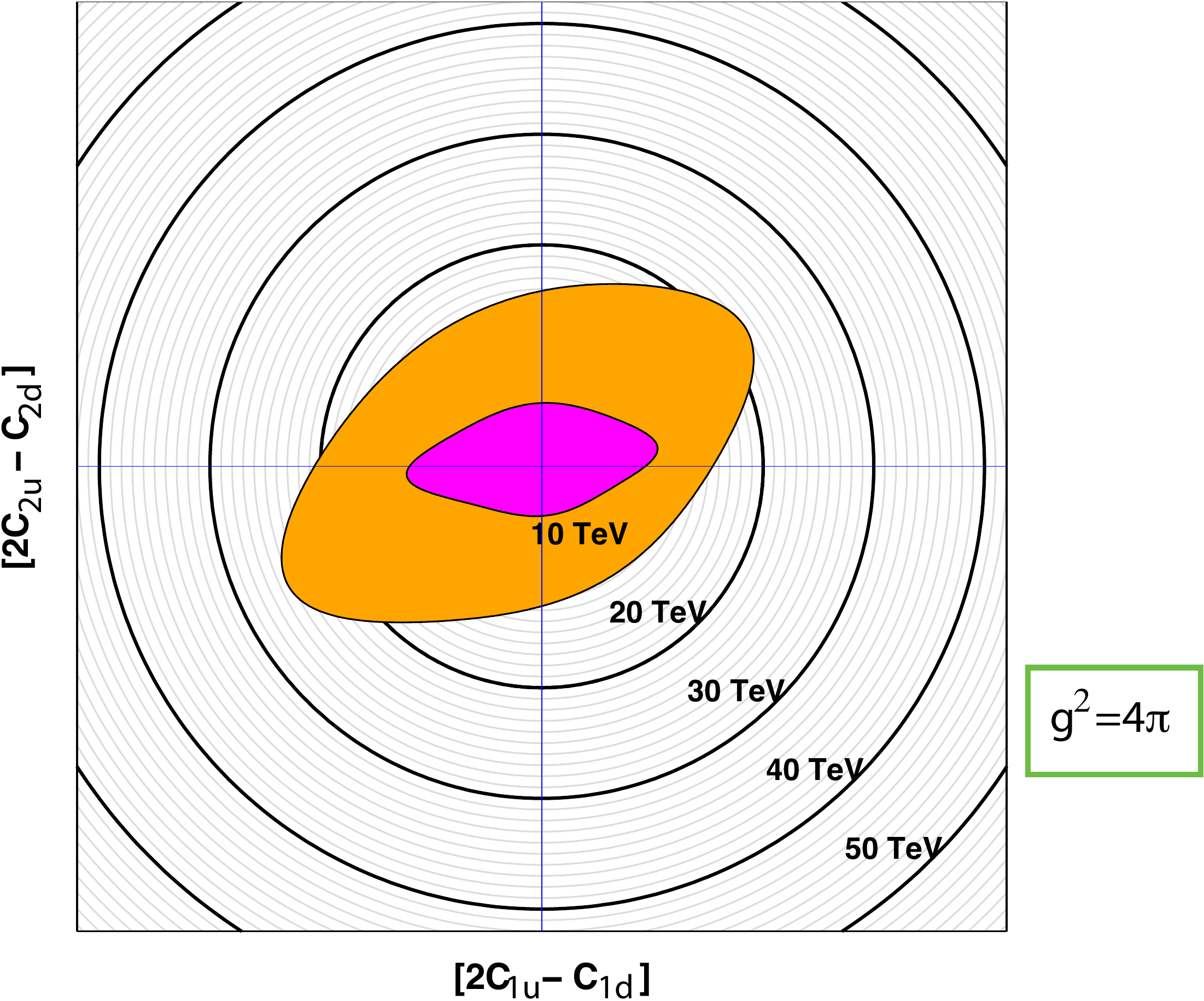}
     \caption{Left panel: Projected uncertainties for $A_{PV}$ in percent as a 
function of kinematics.  The green points are from 11 GeV beam and the 
yellow points are from 6.6 GeV beam.
Right panel: Projected mass limits for composite models. The purple 
ellipse shows the mass limits from the 6 GeV PVDIS results in 
combination with the
published $C_{1q}$ results from PVES (including QWeak) and atomic parity 
violation experiments. The orange ellipse shows the mass limits from the 
SoLID-PVDIS projection in combination with the same projected $C_{1q}$ 
results including the estimated sensitivity from the full Qweak data 
sample.}
\label{fig:PV2}
   \end{center}
\end{figure}

It is important to improve the measurement of the $C_{2q}$ to the level of the $C_{1q}$.  This requires a facility with both large acceptance and high luminosity, which leads to the idea of SoLID~\cite{Souder:2007zze,Souder:2005tz}.  We have developed a preliminary conceptual design of a suitable spectrometer, SoLID, which can reach this goal.  The apparatus can measure $A_{PV}$ for about 20 kinematic points shown in the left panel of Fig.~\ref{fig:PV2} with $x>0.4$ and a range of $Q^2$ with a statistical precision of about 0.5\%.  The SoLID spectrometer is part of the parity-violation program at JLab cited in Recommendation I of the 2007 NSAC Long Range Plan.  It is part of the SM initiative listed on p. 76 and described in more details on p. 87 of the above document. 

One way to characterize and compare the reach of different experiments is to quote mass limits within composite models~\cite{Eichten:1983hw}, where the couplings are on the order of $4\pi/\Lambda^2$, where $\Lambda$ is the compositeness mass scale.  Such limits for the JLab PVDIS collaboration and the SoLID PVDIS experiment~\cite{Erler:2014fqa} are shown in the right panel of Fig.~\ref{fig:PV2}.  The mass limits are on the scale probed by the LHC, and have the unique feature that they probe a very specific helicity structure.

Charge symmetry violation (CSV) in the PDF's is an important possible QCD effect that may be large enough to explain the apparent inconsistency of the NuTeV experiment~\cite{Zeller:2001hh} with the SM~\cite{Londergan:2003ij,Londergan:2009kj}. The PVDIS measurements in a broad kinematic region of the SoLID program on a deuterium target will make a first precision determination of the CSV effect in deuterium at the quark level.
Another interesting possibility is an additional modification of the PDF's in heavier nuclei, the 
isovector EMC effect~\cite{Cloet:2012td}  that causes a ``pseudo'' CSV in nuclei with more neutrons than protons.  This effect can be observed by using a $^{48}$Ca target with the SoLID apparatus.
By using a proton target, the SoLID apparatus can measure the $d/u$ ratio in the proton in a manner totally free from nuclear effects.  The data will be complementary to proposed 
experiments at JLab including 
one using mirror $^3$H and $^3$He nuclei to minimize nuclear effects in extracting ${F^n_2}/{F^p_2}$~\cite{Arrington:2011qt}, and the BONUS 
experiment~\cite{Baillie:2011za} by tagging slow moving proton to minimize nuclear effects in extracting $F^n_2$.  Another important possible effect is
$Q^2$-dependence due to higher twist (HT) effects~\cite{Mantry:2010ki}.  In the deuteron, $A_{PV}$
has the special feature that HT can only be due to quark-quark correlations; all other diagrams cancel in the ratio.

\section{ Threshold Electroproduction of the $J/\Psi$ on a Nucleon}
While significant progress has been achieved in exploring QCD in its perturbative regime, much remains to be understood in the strong regime where the theory is hardly tractable. Lattice QCD, offers real hope for serious  progress in unraveling the structure of nucleons. However, given the approximations required to mitigate the limitation of existing computing power for full fledged {\it ab initio} calculations of key observables, a hand in hand collaboration between experimental measurements and calculations is important. 

A natural and basic question to ask in nuclear physics is, how is the mass of the nucleon shared among its constituents and their interactions? In two important  papers Ji~\cite{Ji:1994av,Ji:1995sv} answered the question by providing what he called ``A QCD analysis of the mass structure of the nucleon''. Using the energy-momentum tensor in QCD it was shown that one can partition the mass of the nucleon among four terms identified as the kinetic and potential energy of the quarks, the  kinetic and potential energy of the gluons, the quarks masses and the conformal (trace) anomaly. The wealth of deep inelastic scattering data  was used to estimate the first three terms at a scale of 1 GeV$^2$, while the fourth (conformal anomaly)  term was estimated, assuming the nucleon mass sum rule, to contribute roughly 20\% to the nucleon mass, and this is quite significant. It is worth noting that this term is scale independent which implies that  as we probe the nucleon with higher resolution (Q$^2$ or higher energy scale) the partition of the mass between quark and gluon energies is modified while the conformal anomaly piece remains the same. This first indirect estimation of the conformal anomaly contribution  to the nucleon mass  was an important step forward in understanding the low energy regime of the gluonic structure of the nucleon and QCD, however, neither direct measurements nor lattice calculations of this term have been attempted yet.

\begin{figure}[H]
\begin{center}
\includegraphics[scale=0.5]{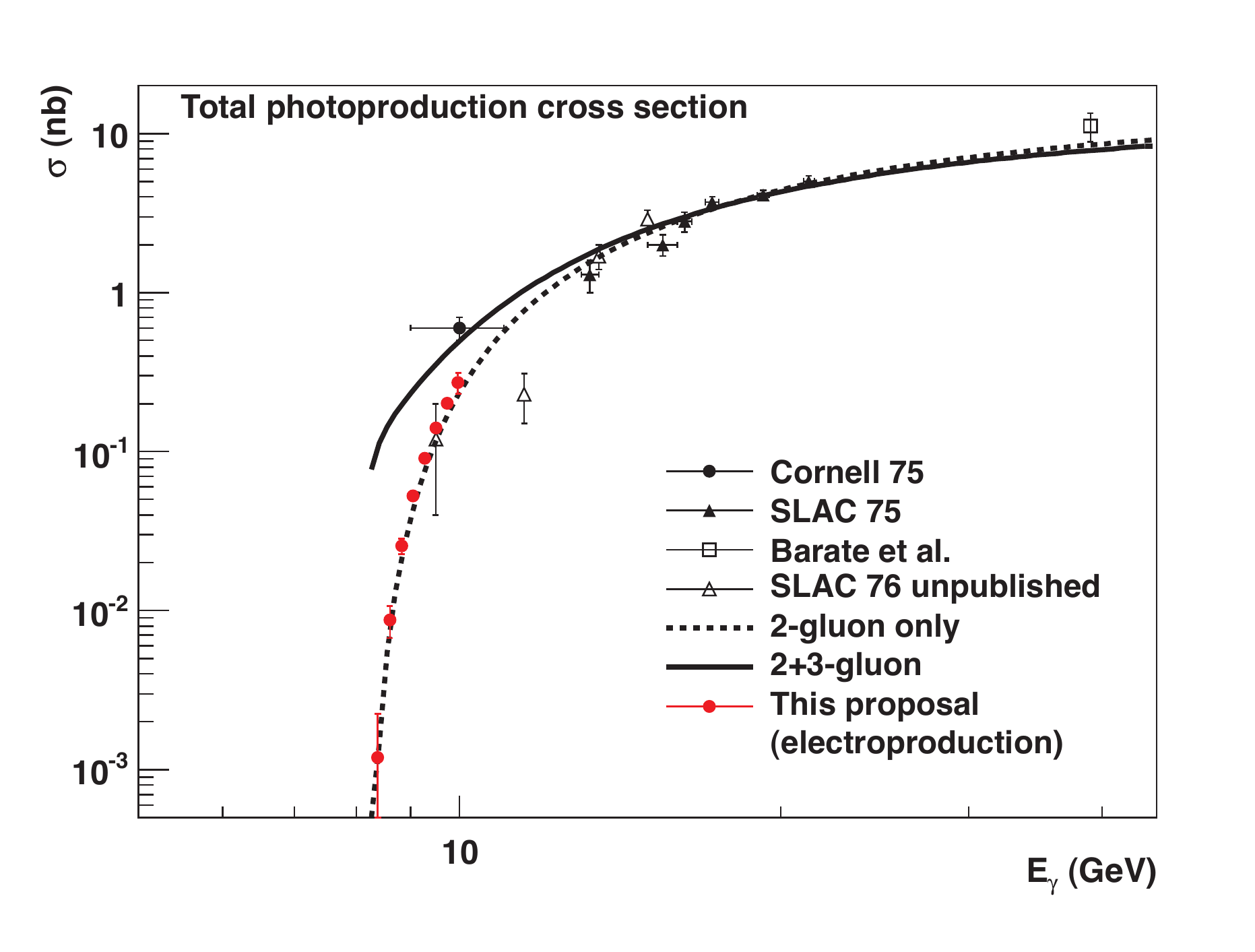}
\caption{Total  $J/\psi$  photoproduction cross section versus photon energy in the threshold region, except for the proposed electroproduction measurement (filled red circles) where the data are plotted as a function of the equivalent photon energy.} 
\label{psi-xsec-w-dep}
\end{center}
\end{figure}

With the 12 GeV upgrade at Jefferson Lab combined with the large acceptance of the SoLID  spectrometer~\cite{pCDR}, designed to operate in a high luminosity environment,  there is a unique opportunity to make a measurement directly sensitive to the anomaly contribution to the $J/\psi$-nucleon interaction at low energy~ \cite{Kharzeev:1995ij,Kharzeev:1998bz} which will be achieved by measuring the exclusive electroproduction of $J/\psi$ near threshold on a proton. With minimal reconfiguration of the semi-inclusive deep inelastic scattering (SIDIS) set up of SoLID, we can perform measurements of the differential electroproduction cross sections near threshold as a function of the four-momentum transfer to the nucleon ($|t-t_{min}|$). From these measurements, the total virtual-photon absorption cross section is evaluated in the region close to the threshold of $J/\psi$ production (see Fig.~\ref{psi-xsec-w-dep}). The strength of the conformal anomaly contribution to the low energy $J/\psi$-nucleon interaction is reflected in the rate with which the total cross section rises (or falls), in the threshold region, as a function of photon energy. Models of the $J/\psi$ production mechanism and its low energy interaction with the nucleon have also been proposed based on two or three gluon exchanges~\cite{Brodsky:2000zc}, they can be tested too.

Lattice QCD calculation of the anomaly contribution to the nucleon mass in tandem with the proposed measurement will be available in time to fully gauge our understanding of the nucleon mass from QCD.

Beyond accessing the trace anomaly the measurement of electroproduction of the $J/\psi$ near threshold opens a host of important questions related to the gluonic structure of the nucleon. What is the strength of the  $J/\psi$-nucleon  interaction near threshold? It is unique in nuclear physics that one can investigate an interaction between two color neutral objects, the nucleon and the $J/\psi$ that is purely gluonic.  In this case the interaction force can be considered to be a color Van der Waals force in analogy to atomic physics. Furthermore, at low energy we do not fully understand the production mechanism of quarkonium ($J/\psi$, $\Upsilon$) on a nucleon.


The $J/\psi$ physics program at threshold using SoLID is unique in its important impact on our understanding of the low energy structure of the nucleon. From understanding where a large fraction of the nucleon mass originates from (conformal anomaly) to exploring the existence of ``color Van der Waals forces'' as well as the production mechanism of the $J/\psi$, SoLID offers the best tool and promise to reach this physics. 

\section{Overview of SoLID Instrumentation}

The SoLID (Solenoidal Large Intensity Device) project will develop a large acceptance spectrometer/detector system capable of handling very high rates. 
It is designed to satisfy the requirements of five approved highly rated experiments, that require both high luminosity and large acceptance, to exploit 
the full potential of the Jefferson Lab 12 GeV upgrade.
 The base equipment composing the SoLID project includes two configurations: the ``SIDIS\&J/$\psi$'' configuration and the ``PVDIS'' configuration.  
Although the geometrical layouts for the detectors are not the same in the two configurations, most components of the following list of the 
SoLID base equipment are in common:
\begin{enumerate}
    \item{A solenoidal magnet with a power supply and cryogenic system, now identified as the CLEO-II magnet. With some modifications as described in the pCDR\cite{pCDR}, this magnet meets the experimental requirements. 
}
     \item{GEM detectors for tracking: These are planned to be provided by a SoLID Chinese Collaboration. Five Chinese institutions (USTC, CIAE, Tsinghua, Lanzhou and IMP), in collaboration with US groups (UVa and Temple), have committed to perform R\&D and plan to apply for funding from the Chinese funding agencies to construct the GEMs for the SoLID project.}
    \item{An electromagnetic calorimeter for electron identification. (In the SIDIS configuration, it is separated into two sectors, a forward sector and a large-angle sector).}
    \item{A light gas Cherenkov detector for electron identification.}
\item{A heavy gas Cherenkov detector for pion (hadron) identification. This is for the SIDIS configuration only.}
     \item{ A MRPC (Multi-Gap Resistive Plate Chamber) detector serving as a time-of-flight (TOF) detector for pion (hadron) identification. 
This is for the SIDIS configuration only. The Chinese groups (Tsinghua and USTC), in 
collaboration with US groups (Duke and Rutgers), have agreed to perform R\&D and apply for funding to construct the required MRPC detector for the SoLID project.}
\item{A set of baffles to reduce background. This is for the PVDIS configuration only.}
    \item{A data acquisition system (DAQ) with online farm capability.}
    \item{Supporting structures for the magnet and the detectors.}
   \item{Requisite Hall A infrastructure to accommodate the functioning of the above.} 
    \end{enumerate}

 The five approved experiments in the SoLID research program would require the SoLID base equipment, as well as the components outside the base equipment of the SoLID project. 
The following lists such additional equipment that is either standard and existing at JLab or that will be available for experiments planned before the SoLID experiments: 
     \begin{enumerate}
\item{For SIDIS transversely, and longitudinally polarized $^3$He: The existing polarized $^3$He target with performance already achieved from the 6~GeV experiment is required.}
\item{ For $J/\psi$ the standard cryogenic LH2 target system with modification in layout is required. }
\item{ For PVDIS: A Compton polarimeter and a super-conducting Moller polarimeter (both also required by the MOLLER project and to be employed for the PREX experiment also) are assumed to be available.}
    \end{enumerate}
 The following items will be required for specific experiments:
 \begin{enumerate}
     \item{ For PVDIS:  a custom, high-power cryotarget is required. Cryogenic cooling capability for the cryotarget (ESR2) is assumed to be available (required by the MOLLER project). }
     \item{ For the SIDIS-proton experiment: a transversely polarized proton target is needed. An initial study has been performed by Oxford which 
concluded that such a target with the specifications satisfying the SoLID experiment requirements is feasible.}
 \end{enumerate}

\section{Current Status}
The SoLID spectrometer was initially proposed in 2009 and approved in 2010 for two experiments:
SIDIS experiment E12-10-006) and the PVDIS experiment E12-10-007. Both experiments aim to achieve high precision which 
require very high statistics. A spectrometer/detector system with a large acceptance 
and also able to handle high luminosity is required. Therefore SoLID is designed to 
have a large solid angle and broad momentum acceptance and can handle luminosity
up to $10^{39}$s$^{-1}$cm$^{-2}$ with a baffle system in the PVDIS configuration and  $10^{37}$s$^{-1}$cm$^{-2}$ without 
a baffle system in the SIDIS\&$J/\psi$ configuration. With these unique features, SoLID is ideal for inclusive and semi-inclusive
DIS experiments and is also good for measurements of certain exclusive reactions.
The SoLID base equipment consists of a solenoid magnet (CLEOII magnet), tracking detectors (GEMs), electron PID detectors
(electromagnetic calorimeter and light gas C\v{e}renkov detector) and hadron PID detectors (MRPC, heavy gas C\v{e}renkov
and EC), DAQ system, supporting structure and infrastruture needed for the spectrometer. 
Leveraging the unique capabilities of SoLID, currently, there are
five highly rated (four ``A'' ratings and one ``A$^-$'') experiments approved using SoLID, along with two 
``parasitic'' experiments. 

The conceptual design has gone through many iterations, including  
various case studies, detailed simulations, pre-R\&D testings and a number of internal reviews.
Of the various internal reviews, it is worth mentioning the two brainstorming sessions in September 2011 and January 2012, organized 
by the JLab physics division, and the dry-run review in June 2012 with external experts. These reviews helped greatly in optimizing, improving and finalizing the conceptual design.
Detailed simulations with realistic background (including neutron backgrounds) and pre-R\&D activities focusing on the major challenges
 have significantly improved the reliability of the conceptual design. The JLab Hall A engineering group, working with the SoLID collaboration, performed studies on the modification of the CLEO-II magnet to satisfy the SoLID needs. The JLab management has formally requested the CLEO-II magnet and was approved by the management from the CLEO side. A recent site visit confirmed that the magnet is preserved in excellent shape. A plan for the extraction and transportation to JLab is in place.  
The final version of the pre-CDR\cite{pCDR} was submitted to JLab management in July 2014.

\section{Summary}

The SoLID spectrometer will be critical in order to meet the major 
challenges and opportunities enabled by the JLab 12 GeV upgrade. 
One key issue is thorough understanding of the three-dimensional structure of the nucleon in the valence region to uncover the rich QCD dynamics and discover new phenomena. Only detailed and systematic measurements of the TMD's using SoLID with the combination of high luminosity and large acceptance, giving very high precision 
in a large number of kinematic bins for both the neutron ($^3$He) and the proton, can deliver the required 
information. 
The existing equipment requires averaging over larger 
kinematic regions which will likely miss many of the key details.

JLab also has had an excellent tradition in electroweak physics, and the 
SoLID spectrometer can continue the tradition, providing a test of the 
SM in a new region of parameter space and also address specific issues 
in nucleon structure including charge symmetry violation and higher twist effects due to 
di-quarks.  Again, the key to this program is high statistics in many kinematic 
bins.  With the presently available facilities, any measurement would be 
unable to untangle these effects.  
 In addition, the $J/\Psi$ physics program at threshold using SoLID is unique in its important impact on our understanding of the gluonic contributions in the strong regime of QCD. This challenging process again requires 
high luminosity and also the ability to measure several particles in 
coincidence. 

The SoLID spectrometer is able to achieve the remarkable performance 
required by the above physics by using a number of new developments in 
instrumentation, including large area GEM tracking detectors, new PMT's 
that can operate for a gas Cerenkov in a moderate magnetic field, and 
pipeline electronics, recently developed for JLab Hall D, that can handle the 
high rates and deal efficiently with dead-time and pileup problems.

In summary, the SoLID spectrometer is a remarkably flexible device that 
can greatly enhance the physics output of the JLab upgrade in a number 
of exciting areas.  In addition, it will provide a proving ground for 
some of the important technologies that will be crucial for moving 
forward towards a future Electron Ion Collider detector.

\section{Acknowledgement}

We thank K. Allada, T. Averett, R. Beminwattha, A. Camsonne, A. Courtoy, W. Deconinck, 
R. Ent, K. Hafidi, J. Huang, X.D. Jiang, C. Keppel,
X.M. Li, N. Liyanage, R.D. McKeown, M. Meziane, H. Montgomery,
B. Pasquini, M. Paolone, J.C. Peng, A. Prokudin, X. Qian,
Y. Qiang, P. Reimer, S. Riordan, C. Roberts, N. Sparveris,
A.W. Thomas, Z. Xiao,
W.B. Yan, H. Yao, Z. Ye, J. Zhang, Z.W. Zhao, and X. Zheng for their help in the preparation of this white paper.
This material is based upon work supported 
in part by US Department of Energy, Office of Science, Office of Nuclear Physics under contract numbers DE-AC05-06OR23177 (JLab),
DE-FG02-03ER41231 (Duke), DE-FG02-94ER40844 (Temple), DE-FG02-96ER40988 (Stony Brook U.), and DE-FG02-84ER40146 (Syracuse U.).

\end{document}